\documentclass[12pt, ams, nofootinbib]{revtex4-1}
\usepackage{amsmath}
\usepackage{amsfonts}
\usepackage{amsthm}
\usepackage{amssymb}
\usepackage{graphicx}
\usepackage{epsfig}
\usepackage{hyperref}
\usepackage{amscd}
\usepackage{xcolor}
\usepackage{enumerate}

\usepackage{frcursive}
\usepackage[T1]{fontenc}

\usepackage[toc]{appendix}

\def\be{\begin{equation}}
\def\ee{\end{equation}}

\theoremstyle{definition}

\theoremstyle{Identity}

\begin{document}
\preprint{APS/123-QED}
\title{Fermi-normal coordinates for the Newtonian approximation of gravity}

\author{Antonio  C. Guti\'errez-Pi\~{n}eres}
\email{acgutier@uis.edu.co}
\affiliation{Escuela de F\'\i sica, Universidad Industrial de Santander, CP 680002,  Bucaramanga, Colombia}
\affiliation{Instituto de Ciencias Nucleares, Universidad Nacional Aut\'onoma de M\'exico,
 AP 70543,  M\'exico, CDMX 04510, M\'exico}

 \begin{abstract} 
In this work, we compute the metric corresponding to a static and spherically symmetric mass distribution 
in the general relativistic weak field approximation to quadratic order in Fermi-normal coordinates surrounding a radial geodesic. 
To construct a geodesic and a convenient tetrad transported along it, we first introduce a general metric, use the 
Cartan formalism of differential forms, and then specialize the space-time by considering the nearly Newtonian metric. 
This procedure simplifies the calculations significantly, and the expression for the radial geodesic admits a simple form.
We conclude that in quadratic order, the effects of a Schwarzschild gravitational field measured locally by a freely falling 
observer equals the measured by an observer in similar conditions in the presence of a Newtonian approximation of gravitation.
\end{abstract}


\maketitle

\section{Introduction}
The local flatness theorem ensures that finding a local Lorentz frame at a point $P$ in a given space-time is always 
possible. That means that, for a point $P$ of space-time, it is always possible to find a coordinate system in which the 
metric tensor corresponds to the Minkowski tensor with zero Christoffel symbols. Therefore, free-falling observers 
do not see any effect of gravity in their immediate vicinity \cite{misner2017gravitation, poisson2004relativist}. 

Pioneering ideas about this theorem probably came from the definitions of the Riemann-normal coordinates 
of a point $P$ in space-time. In such coordinates of the point $P$, the metric tensor corresponds to the Minkowski 
metric, with the first derivative vanishing and the second derivatives expressed only in terms of the Riemann tensor.
 More general coordinate systems exist where a single point admits extension to an arbitrary curve, leading to the 
 metric orthogonal and few nonzero Christoffel symbols. These coordinate systems are essential in numerous physical
  situations \cite{synge1960relativity}.

Under Fermi transport, the fourth member of an orthonormal tetrad remains tangent on a general curve. The other members
 of the tetrad provide a spatial triad forming a reference frame for an observer whose word line is precisely the general curve 
 in question. On that orthonormal tetrad, it is possible to construct a reference frame at which the metric is orthogonal and 
 with nonzero Christoffel symbols along the observer word line. It is common to refer to this tetrad as the Fermi reference 
 frame or simply as the Fermi coordinates. This frame gives us the correct relativistic generalization of the Newtonian concept 
 of a non-rotating frame.  

An exciting and particular case of the Fermi coordinates occurs when the curve in question is a geodesic. In this case, 
the Fermi transport of the orthonormal tetrad coincides with its parallel transport. If the operating coordinates satisfy 
the additional condition that the Christoffel symbols vanish along the geodesic, the resulting coordinates are called 
Fermi-normal coordinates. Manasse and Misner \cite{manasse1963fermi, manasse1963distortion} primarily used 
these coordinates to determine the lowest-order effects of a Schwarzschild gravitational field, which can be measured 
locally by a freely falling observer.

 For as long as the theory of General Relativity has existed, there has been debate about the corresponding question 
of whether and to what extent Newton's theory of gravitation can be considered a special case or, at least, a limiting 
situation of General Relativity. Much work has been dedicated to the study of Schwarzschild spacetime, and very special 
coordinates were used to establish its link with Newton's theory of gravitation. Likewise, there are numerous works dedicated 
to clarifying the transition from the Einsteinian gravitational potential to the Newtonian, and vice versa.
The main interest of this work focuses on reinforcing existing knowledge, emphasizing the value of new tools and perspectives 
to understand established principles.

  In this work, we ask ourselves how a freely falling observer determines the quadratic order effects of a Newtonian 
  approximation of gravitation, and we review the well-understood principle of local equivalence between the effects 
  of a Schwarzschild gravitational field and the Newtonian gravitational approximation. To answer the question, 
  we construct a reference geodesic curve and transport a Fermi-normal frame along it. Following the ideas in 
  the early paper by Manasse and Misner \cite{manasse1963fermi},  Section \ref{sect:A} presents the geometric construction 
  of the Fermi-normal coordinates and expresses the metric near the geodesic
  as to quadratic order in these coordinates. 
 
The main advantage of using the normal Fermi coordinates is that, on the basis of these coordinates, 
an observer in free fall can quickly determine locally the Newtonian approximation of gravitation.
  To obtain the metric to quadratic order in Fermi-normal coordinates, we must express its coefficients as functions
 of the components of the Riemann curvature tensor evaluated on the geodesic used to construct these coordinates. 
 Because of this, we need to know the geodesic explicitly and determine the orthonormal tetrad, which is parallel 
 transported on the reference geodesic. Usually, this is more complex than it may seem.

We use local orthonormal tetrads and the Cartan formalism of the differential forms to simplify the calculations and clarify 
the procedure employed. From the physical point of view, this procedure is promising; in fact, a local orthonormal tetrad 
is the most straightforward and natural choice for an observer to perform local measurements of time, space, and gravity. 
Furthermore, once one chooses a local orthonormal tetrad, all the quantities related to this frame are invariant with respect 
to coordinate transformations.

From the mathematical point of view, the Cartan formalism of the differential forms is significantly helpful and allows 
us to represent the curvature Riemann tensor in its most straightforward manner. Additionally, it leads an observer 
placed on a local orthonormal tetrad transported parallelly on a reference geodesic, which looks like a right-line as 
viewed in such local tetrad. Evidently, the Cartan formalism of differential forms provides 
a streamlined and insightful approach to constructing geodesics.

In Section \ref{sect:B}, we introduce a general metric, express it in an orthonormal tetrad,  use the Cartan formalism 
of differential forms, and finally, specialize the space-time by considering the nearly Newtonian metric. 
 As far as we know, the precedent idea has yet to be explicitly discussed; equivalent procedures engage a mathematical 
process implying laborious techniques to obtain the reference geodesic. In Section \ref{sect:C}, we specialize the space-time by considering 
the nearly Newtonian metric. Section \ref{sect:D} presents some concluding remarks and areas for further research.

\section{Geometric construction of Fermi-normal coordinates}\label{sect:A}
In this section, we briefly illustrate the procedure for constructing Fermi-normal coordinates. To do so, we first 
consider a given space-time with a metric expressed in arbitrary coordinates $x^{\tilde \alpha}$
                  \footnote{Fermi-normal coordinates will labelled as $x^{\alpha}$.  We shall use the  conventions 
                                 according to which  Greek indices $\alpha, \beta, \dots,$  and  first Latin indices  $a, b, \dots,$ 
                                 run over $0,\dots,3$,  while mid-Latin indices as  $i, j, \dots,$ run over 1,\dots, 3.} .
Then, we choose a point $O$ as the origin and assume a time-like geodesic 
$\gamma$, which starts at this point. To describe the geodesic, we let $t$ be a proper time along it so that 
the point $O$ is given by taking $t=0$, that is, $O=\gamma(0)$.

We erect an orthonormal basis of vectors $\mathbf{e_{0}}, \mathbf{e_{1}}, \mathbf{e_{2}}$ and $\mathbf{e_{3}}$ to fix 
the coordinates axis at $O$ with tangent 
                                               $\mathbf{e_{0}} = \mathbf{e_{0}}(0)$.                                                
Because $\gamma$ is a geodesic, its tangent at two arbitrary points will relate by parallel displacement along 
$\gamma$. Therefore, the vector $\mathbf{e_{0}}(t)$ will be tangent to the geodesic in any point $P = \gamma(t)$ 
on it.  Similarly, the other vectors $\mathbf{e_{i}}(0)$ of the tetrad are parallel displaced to get $\mathbf{e_{i}}(t)$ at 
any  arbitrary point $P$. We can assume for simplicity that $\mathbf{e_{0}}(t)$ is a time-like vector and 
$\mathbf{e_{i}}(t)$  are space-like vectors.

We can choose a time-like vector  
                                                ${\mathbf u(0)} = {\mathbf e_{0}}(0)$
  and a vector 
                                                ${\mathbf v(0)} = v^{i}{\mathbf e_{i}}(0)$
orthonormal to it, i. e.
                                                  ${\mathbf u(0)} \cdot {\mathbf v(0)} =0 $.  
In terms of the arbitrary coordinates, these vectors admit the relation
         \begin{align}
           u^{\tilde \mu}   & =  (\mathbf{e_{\alpha}})^{\tilde \mu} u^{\alpha} = (\mathbf{e_{0}})^{\tilde \mu} \ , 
            \label{eq:orthonormal1}
             \end{align}
          \begin{align}                    
           v^{\tilde \mu}   & =  (\mathbf{e_{\alpha}})^{\tilde \mu} v^{\alpha} = (\mathbf{e_{i}})^{\tilde \mu}  v^{i} \ ,
           \label{eq:orthonormal2}
          \end{align}
where we chose, conveniently,  $v^0 =0, $  and $u^{\alpha} = \delta^{\alpha}_0$, as well as
         \begin{align}
           \eta_{ \mu \nu} & = (\mathbf{e_{\mu}})^{ \tilde \alpha} (\mathbf{e_{\nu}})^{ \tilde \beta}
                                          g_{{ \tilde \alpha} { \tilde \beta}} \ ,
           \label{eq:metrics-relation1}
            \end{align}
everywhere on $\gamma$. Here $g_{{ \tilde \alpha} { \tilde \beta}}$ is the metric in arbitrary coordinates, and 
                                                       $ \eta_{\mu\nu}  = {\text diag}(-1, 1, 1, 1)$
is the Minkowski metric.

Let us now consider a space-like geodesic $\beta$ originating at a point $P$ on $\gamma$, at which $t = t_P$. 
Then, given $x^{\alpha}$, we construct the vector  ${\mathbf v(t)} = v^{i}{\mathbf e_{i}}(t)$ tangent to $\beta$ at  
point $P$. Naturally, we do
                                \begin{align}
                                  v^{i} = \frac{x^{i} }{s} \ , \quad  
                                    s^2 \equiv (x^1)^2   + (x^2)^2  + (x^3)^2  \ ,
                                    \label{eq:tangent0}
                                 \end{align}
to normalize ${\mathbf v}$ properly.  We can construct a family of space-like geodesics $\beta(t, v^{i})$ orthogonal 
to $\gamma$ at $P$ by choosing  different $v^{i}$.  This family of geodesics has proper distance $s$ along $\beta$; 
we set  $s=0$  at $P$. Then,  a unique geodesic intersects $\gamma$ orthogonally at $P$, is tangent to the vector ${\mathbf v}$, 
 and passes through a point $Q$. To specify the point $Q$ in Fermi-normal coordinates, we then write 
                                              \begin{align}
                                               Q= \beta(t, v^{i}, s) \ , 
                                                \label{eq:FermiCoor}
                                                \end{align}
 with $t$ denoting proper time at the intersection point,  $v^{i}$ the components of the vector ${\mathbf v(t)}$ at that point, 
 and $s$ the proper distance from $P$ to $Q$.

Equation $(\ref{eq:FermiCoor})$ determines  a point of a geodesic in Fermi-normal coordinates; we can rewrite it in 
some arbitrary coordinates 
                                               $x^{\tilde \alpha} = (Q)^{\tilde \alpha} $ 
as 
                                                \begin{align}
                                                x^{\tilde \alpha} = x^{\tilde \alpha}(t, v^{i}, s) \ .å
                                                 \label{eq:ArbitraryCoor}
                                                 \end{align}
It is not difficult to demonstrate that, given an arbitrary parameter $\kappa$, 
                                                 $ x^{\tilde \alpha}(t, s v^{i}, \kappa) $
and
                                                 $x^{\tilde \alpha}(t, v^{i}, s \kappa) $
 satisfy the same geodesic equation and initial condition. Hence, the uniqueness theorem  for solutions of 
 differential equations give us 
                                                 \begin{align}
                                                 x^{\tilde \alpha}(t, v^{i}, s \kappa) = x^{\tilde \alpha}(t,  s v^{i}, \kappa)  \ , 
                                                 \label{eq:uniqueness}
                                                 \end{align}
and,  as a special case,  we find
                                                 \begin{align}
                                                 x^{\tilde \alpha}(t, v^{i}, s) = x^{\tilde \alpha}(t,  s v^{i}, 1) 
                                                                                          \equiv  x^{\tilde \alpha}(x^{\alpha}) \ , 
                                                 \label{eq:uniquenessV2}
                                                 \end{align}
which gives a transformation rule between  $x^{\tilde \alpha}$ and the Fermi-normal coordinates.

Finally, the tangent vector to the geodesics ${\beta}(t, v^{i}, s)$ in the coordinates $x^{\tilde \alpha}$ is 
                                                 \begin{align}
                                                   v^{\tilde \alpha} = \frac{ \partial x^{\tilde \alpha}}{\partial s }{\bigg |}_{t, v^{i}}  
                                                    \equiv     \frac{ d x^{\tilde \alpha}}{ds } \ .
                                                    \label{eq:tangentt}
                                                     \end{align}   
The notation indicates that we take the derivative with respect to $s$ while keeping $t$ and $v^{i}$ fixed.  Accordingly, 
this vector calculated in the geodesic   $\gamma$ reveals that
                                                   \begin{align}
                                                     v^{\tilde \alpha}   {\big |}_{\gamma}  = 
                                                    \frac{ d x^{\tilde \alpha}}{ds }{\bigg |}_{\gamma} 
                                                     =   \frac{ \partial x^{\tilde \alpha}}{\partial x^{i} } {\bigg |}_{s=0}  \frac{dx^{i}}{ds}
                                                     =  \frac{ \partial x^{\tilde \alpha}}{\partial x^{i} }{\bigg |}_{s=0} v^{i}  \ ,
                                                    \label{eq:thetrad-relation}
                                                     \end{align}   
where we used the relation $x^{i} = s v^{i}$ given by Equation (\ref{eq:tangent0}). Thus, a direct comparison between (\ref{eq:orthonormal2}) and (\ref{eq:thetrad-relation})  gives
                                                  \begin{align}
                                                  (\mathbf{e_{i}})^{\tilde \alpha}  =  \frac{ \partial x^{\tilde \alpha}}{\partial x^{i} } 
                                                                                                      {\bigg |}_{\gamma} \ .
                                                   \label{eq:orthonormal3}
                                                     \end{align}   
Additionally,  from (\ref{eq:orthonormal1}) we know that
                                                   \begin{align}
                                                    (\mathbf{e_{0}})^{\tilde \alpha}  =  \frac{ \partial x^{\tilde \alpha}}{\partial t} 
                                                                                                          {\bigg |}_{\gamma} \ .
                                                    \label{eq:orthonormal4}
                                                     \end{align}  
 Therefore, from (\ref{eq:orthonormal3}) and (\ref{eq:orthonormal4}) we conclude that along the geodesic $\gamma$,
                                                 \begin{align}
                                                  ({\mathbf e_{\mu} })^{\tilde \alpha} 
                                                                      = \frac{ \partial x^{\tilde \alpha}}{\partial x^{\mu} }{\bigg |}_{\gamma} \ ,
                                                    \label{eq:orthonormalF}
                                                     \end{align} 
or, equivalently,                                                   
                                                   \begin{align}
                                                   {\mathbf e_{\mu}}(t) = \frac{ \partial }{\partial x^{\mu} } {\bigg |}_{\gamma} \ ,
                                                    \label{eq:orthonormalF2}
                                                     \end{align} 
is the orthonormal tetrad on the geodesic $\gamma$ in the Fermi-normal coordinates.

Furthermore, we note that the relation between metric components in the Fermi-normal coordinate $x^{\alpha}$ and t
he components of the metric in the arbitrary coordinates $x^{\tilde \alpha}$ is given by 
         \begin{align}
           g_{ \mu \nu} & = \frac{\partial  x^{ \tilde \alpha} }{\partial x^{\mu} } \frac{\partial x^{ \tilde \beta} }{\partial x^{\nu}} 
                                         \, g_{{ \tilde \alpha} { \tilde \beta}} \ .
           \label{eq:metrics-relation2}
            \end{align}
Thus, by  evaluating all the quantities on the geodesic  and using  (\ref{eq:orthonormalF}), we get
         \begin{align}
          g_{ \mu \nu}{\bigg |}_{\gamma} & =      ({\mathbf e_{\mu} })^{\tilde \alpha}  ({\mathbf e_{\nu} })^{\tilde \beta} 
                                                                      \, g_{{ \tilde \alpha} { \tilde \beta}} \ .
           \label{eq:metrics-relation3}
            \end{align}
Hence, a direct comparison between (\ref{eq:metrics-relation1}) and (\ref{eq:metrics-relation3}) gives
         \begin{align}
          g_{ \mu \nu}{\bigg |}_{\gamma} & =    {\mathbf e_{\mu} } \cdot {\mathbf e_{\nu} } = \eta_{ \mu \nu}\ .
           \label{eq:metrics-relation4}
            \end{align}
The last result shows that, in the Fermi-normal coordinates, the metric is Minkowski everywhere on the geodesic 
 $\gamma$. This result is consistent with the local-flatness theorem and the specialization of the Fermi original ideas. 
 So, given a geodesic, it is possible to introduce Fermi-normal coordinates $x^{\alpha} =(t, x^{i})$ near this so that 
 the  Christoffel symbols vanish along. Furthermore, to express the metric near the geodesic as 
 \cite{manasse1963fermi}
                      \begin{align}
                        g_{00}   & = -1\ +\   R_{0 l 0 m} {\bigg |}_{\gamma} \, x^{l} x^{m}  \label{eq:F-N-Ma}\ ,\\
                        g_{0 i}   & =  0\  +\  \frac{2}{3}  R_{0  l  i  m}{\bigg |}_{\gamma}\, x^{l} x^{m}\label{eq:F-N-Mb}\ ,\\
                        g_{i j}    & =  \delta_{i j}\ +\  \frac{1}{3} R_{i l j m}{\bigg |}_{\gamma} \, x^{l} x^{m} \label{eq:F-N-Mc}\ .                      
                       \end{align}
Here, $t$ is proper time along the geodesic $\gamma$, on which the spatial coordinates $x^{i}$ vanish, and the 
curvature tensor is evaluated. The dependence of the metric on $t$ is contained entirely in the curvature tensor 
components.  Naturally, the metric corresponds to the Minkowski metric on the geodesic $x^{i} =0$.
\section{The nearly Newtonian space-time}\label{sect:B}
 Linearized theory of gravity is a weak-field limit of general relativity. In this case, the metric coefficients 
 expressed as 
        \begin{align}
        g_{\mu\nu}= \eta_{\mu\nu} + h_{\mu\nu} 
         \end{align}
satisfies a linear approach to the  Einstein field gravitational equations  (See \cite{synge1960relativity}).  Here $\eta_{\mu\nu} = {\text{diag}} (-1,1,1,1)$,
        \begin{align}
         h_{\mu\nu}(t, {\bf x}) \equiv 4\int{ \frac{1}{| {\bf x}  - {\bf x'}|} \ T_{\mu\nu}^{\text{ret}}(t - |{\bf x} -{\bf x'}|,{\bf x'}) \ d^3x' } \  ,
         \end{align}
and ``ret'' means the retarded value. Consequently, the stress-energy components that appear in the linearized field equations
 are the components obtained using special relativity, and, therefore, their conservation does not contain gravitational effects 
\cite{misner2017gravitation} .

An engaging model of relativity occurs when bodies are kept at rest relative to each other by a material medium
 that fills the space between them. In this case, the weak gravitational fields approach allows for a specialization called nearly 
 Newtonian fields and the corresponding space-time, nearly Newtonian space-time. 
 Here, the source of the gravitational field approximates a nearly Newtonian source: $T_{00} \gg |T_{0i}|$, $T_{00} \gg |T_{ij}|$, and 
 velocities slow enough that retardation is negligible. In such case
         \begin{align}
        g_{\mu\nu}= \eta_{\mu\nu} + h_{\mu\nu}   = \eta_{\mu\nu} - \delta_{\mu\nu} \phi \ ,
         \end{align}
where $\delta_{\mu\nu} = {\text{diag}} (1,1,1,1)$ and $\phi$ is a Newtonian potential $\phi \ll1$. In terms of general relativity, the metric 
tensor corresponding to a nearly Newtonian space-time has the following form  (see \cite{misner2017gravitation}, page 470 and \cite{synge1960relativity}, page 206) :
     \begin{align}
     {\cal G} =   - (1 + 2\phi)  dT \otimes dT  +  (1 -  2\phi)  ( dx \otimes dx  + dy \otimes dy
                                                                                         + dz \otimes dz  ) \ .
     \label{eq:NewtAppCart}                                                                                    
      \end{align}
In the next section, we will calculate the metric corresponding to the nearly Newtonian space-time to quadratic order 
in Fermi-normal coordinates surrounding a radial geodesic. To do it, in the current section, we first must calculate the 
curvature tensor of the corresponding space-time in the Fermi-normal coordinates. This procedure requires previously 
computing the curvature of the given space-time in arbitrary coordinates. We initially introduce a general metric, express 
it in an orthonormal tetrad, and use the Cartan formalism of differential forms to emphasize the independence from the 
 coordinates. Finally, specialize the space-time by considering the nearly  Newtonian metric.

Consider a metric ${\cal G} $ in coordinates $y^{a}$  and a set of differential forms 
                          $\vartheta^{\hat \alpha}  \equiv d y^{\hat \alpha}$,  
such that 
     \begin{align}
       {\cal G}  = g_{a b} dy^a \otimes dy^b  =  \eta_{\hat  \alpha  \hat \beta  } \vartheta^{ \hat \alpha} 
                                                                      \otimes \vartheta^{\hat \beta}\ ,
                                                                      \label{eq:GeneralMetric}
      \end{align}
with 
     $\eta_{\hat \alpha \hat \beta}={\rm diag}(-1,1,1,1)$,
and 
     $\vartheta^{\hat \alpha} = ({\mathbf e^{\hat \alpha}})_{a}dy^a$.  The first and second Cartan equations  
     (See \cite{chandrasekhar2002mathematical, misner2017gravitation}) , 
      \begin{align}
                          d\vartheta^{\hat  \alpha} & = - \omega^{\hat  \alpha}_{\ \hat \beta}\wedge \vartheta^{\hat \beta}\ , \\
        \Omega^{\hat  \alpha}_{\ \hat \beta} & = d\omega^{\hat  \alpha}_{\ \hat \beta} 
                                                                     + \omega^{\hat  \alpha}_{ \ \hat \mu} \wedge \omega^{ \hat \mu}_{\ \hat \beta} 
                                                                     =  \frac{1}{2} R^{\hat  \alpha}_{\ \hat \beta   \hat \mu  \hat \nu} 
                                                                          \vartheta^ { \hat \mu} \wedge\vartheta^{ \hat \nu}
          \end{align}
allow us to compute  the Riemann curvature tensor components $R_{ \hat \alpha  \hat \beta \hat \mu \hat \nu}$  
in the local orthonormal frame $\vartheta^{ \hat \alpha}$.  As an additional point, if  we introduce the bivector representation  that defines  
the curvature components $R_{ \hat \alpha  \hat \beta \hat \mu \hat \nu}$ as the 
 components of a $6\times 6$ matrix ${\bf R}_{AB}$ according to the procedure discussed in \cite{gutierrez2019c3, gutierrez2022darmois}, 
 it is possible to calculate the corresponding eigenvalues of the curvature tensor.

 To specialize the metric  (\ref{eq:GeneralMetric}) so that  it corresponds to the nearly Newtonian space-time 
 given by(\ref{eq:NewtAppCart}), 
  let us consider the following metric in spherical coordinates 
                                             $y^{a} = (T, R, \Theta, \Phi)$
in the form
     \begin{align}
     {\cal G} =   - (1 + 2\phi)  dT \otimes dT  +  (1 -  2\phi)  ( dR \otimes dR  + R^2 d\Theta \otimes d\Theta 
                                                                                         + R^2 \sin^2 \Theta d\Phi \otimes d\Phi  ) \ .
     \label{eq:NewtApp}                                                                                    
      \end{align}
Here  $\phi \ll 1$ is the Newtonian potential of a static mass distribution in a not-rotating local frame of reference. 
 As we can see in (\ref{eq:NewtApp}), the $(1 -  2\phi)$  factor multiplies not only the radial component, such as in the Schwarzschild solution 
  of the complete gravitation fields. Conversely, by definition, the $(1 -  2\phi)$ factor also multiplies the angular component of the metric. 
In this work, we limit ourselves to studying  spherically symmetric gravitational configurations,  assuming 
 that $\phi$ depends  on $R$.   
Remarkably, the fact that the $(1 -  2\phi)$ factor multiplies the angular part in general does not break the symmetry of the 
spherical space-time considered if the Newtonian potential depends only on the radial coordinate. 
In previous references, authors present some particular solutions of the Poisson equation determining specific forms of the Newtonian 
gravitational potentials. Each one of these solutions preserves the spherical symmetry of the nearly Newtonian space-time.  
However, as we shall see in the next section, performing a quadratic order of the nearly Newtonian gravitational field breaks the spherical 
symmetry if Fermi-Normal coordinates are employed.

 The components of the orthonormal tetrad are then 
      \begin{align}
       \vartheta^{\hat{0}}  & = \sqrt{1 + 2 \phi} \,dT \ \equiv d \mathcal{T} \ ,\quad
         \vartheta^{\hat{1}} = \sqrt{1 - 2 \phi} \, dR\   \equiv d \mathcal{ R }\ , 
            \label{eq:one-form-a} \\
              \vartheta^{\hat{2}}  &= \sqrt{1  {-} 2 \phi}\, R \, d\Theta \equiv  d\mathcal{\theta} \ ,\quad
              \vartheta^{\hat{3}}  = \sqrt{1  {-}  2 \phi}\,R \sin\Theta \, d\Phi \equiv d \mathcal{\varphi}\ ,
               \label{eq:one-form-b}
               \end{align}
which, in the first-order approximation, leads to the connection 1-form 
          \begin{align}
           \omega^{\hat 1}_{\  \hat 0}  &=   -\phi_R \vartheta^{\hat 0}\ ,  \quad  
	   \omega^{\hat 2}_{\ \hat 3}  = -\frac{1}{R} (1+\phi) \cot{\Theta} \vartheta^{ \hat 3}\ , \nonumber\\
            \omega^{ \hat 1}_{\ \hat 2}   & = -\frac{1}{R} (1+\phi - R \phi_R) \vartheta^{ \hat 2}\ , \quad
             \omega^{ \hat 1}_{\ \hat 3}   = - \frac{1}{R} (1+\phi - R \phi_R) \vartheta^{ \hat 3} \ .
               \end{align}
Moreover, the only non-vanishing components of the curvature 2-form can be expressed up to the first order in $\phi$ as
         \begin{align}
           \Omega^{ \hat 0}_{\ \hat 1}    & =  -  \phi_{RR} \, \vartheta^{ \hat 0} \wedge \vartheta^{ \hat 1}\ , \ \
           \Omega^{ \hat 0}_{\  \hat 2}      =     -  \frac{1}{R}   \phi_R  \,  \vartheta^{ \hat 0} \wedge \vartheta^{ \hat 2}, \ \
            \Omega^{ \hat 0}_{\ \hat 3}      =   -  \frac{1}{R}   \phi_R  \,  \vartheta^{ \hat 0} \wedge \vartheta^{ \hat 3} \ , \\
            \Omega^{ \hat 2}_{\  \hat3}    & =      \frac{2}{R}  \phi_R \vartheta^{ \hat 2} \wedge \vartheta^{ \hat 3} \ , \ \
            \Omega^{ \hat 3}_{\ \hat 1}      =      ( \phi_{RR} + \frac{1}{R}  \phi_{R} )  \vartheta^{ \hat 2} \wedge \vartheta^{ \hat 3}\ , \ \
            \Omega^{ \hat 1}_{\ \hat 2}      =      ( \phi_{RR} + \frac{1}{R}  \phi_{R} )   \vartheta^{ \hat 1} \wedge \vartheta^{ \hat 2} \ .
             \end{align}
It then follows that the only non-zero components  of  the curvature  tensor are 
          \begin{align}
            R_{{ \hat 0} { \hat 1} { \hat 0} { \hat 1}} & = {\bf R}_{{ \hat 1} { \hat 1}} =     \phi_{RR} \, , \ \
            R_{{ \hat 0} { \hat 2} { \hat 0} { \hat 2}}    = {\bf R}_{{ \hat 2} { \hat 2}}  
                                                                            =   R_{{ \hat 0} { \hat 3} { \hat 0} { \hat 3} } 
                                                                           = {\bf R}_{{ \hat 3} { \hat 3}} =    \frac{1}{R}   \phi_R \, 
                                                                                       \label{eq:CurvatureSpherCoord}  
                                                                           ,\\
            R_{{ \hat 2} { \hat 3} { \hat 2} { \hat 3}}& = {\bf R}_{{ \hat 4} { \hat 4}} =   \frac{2}{R}  \phi_R  \, , \ \
            R_{{ \hat 3 }{ \hat 1} { \hat 3} { \hat 1}}   = {\bf R}_{{ \hat 5} { \hat 5}} =  R_{{ \hat 1} { \hat 2} { \hat 1} { \hat  2}}    
                                                                           = {\bf R}_{{ \hat 6} { \hat 6}} =    \phi_{RR} + \frac{1}{R}  \phi_{R}  \, .
            \end{align}
 Consequently, the curvature matrix ${\bf R_{AB}}$ is diagonal with eigenvalues
            \begin{align}
              \lambda_{1}   &  = \phi_{RR}  \ ,  \ \
              \lambda_{2}        =    \lambda_{3}   =    \frac{1}{R} \phi_R\ , \\ 
              \lambda_{4}   &  = \frac{2}{R}  \phi_R\ , \ \
              \lambda_{5}       =     \lambda_{6}  =    \phi_{RR}  +  \frac{1}{R} \phi_R  \ ,
              \label{eq: eigenvs_of_R_inN_V2}
                \end{align}
which satisfies the relationship
             \begin{align}
              \sum_{i=1}^{6}\lambda_{i}= 3\left( \phi_{RR} + \frac{2}{R} \phi_R \right) =  3\ \nabla^2 \phi \ ,
               \end{align}
where $ \nabla^2$ denotes the usual Laplace operator in the spherical coordinates. Furthermore, in Newtonian gravity, the  eigenvalue 
problems of the Riemann curvature give us
              \begin{align}
               \sum_{i=1}^{6}\lambda_{i}= \frac{3\kappa}{2} \rho \,  \Leftrightarrow  \, \nabla^2 \phi = \frac{\kappa}{2} \rho \ .
               \label{poi}
                \end{align}
Here, $\rho$ represents the mass density of the source corresponding to the gravitational. This last result indicates that the approach 
for determining the eigenvalues presented in  \cite{gutierrez2019c3} is compatible 
with the Poisson equation. It then follows that in a region containing no mass,  i.e., $ \rho = 0$,   the gravitational potential 
of a source of mass $M$ is written as $ \phi= - {M}/{R}$.  Hence, the only non-zero components  of  the curvature  tensor are 
          \begin{align}
            R_{{ \hat 0} { \hat 1} { \hat 0} { \hat 1}} & = {\bf R}_{{ \hat 1} { \hat 1}} =   - \frac{2 M}{R} \, , \ \
            R_{{ \hat 0} { \hat 2} { \hat 0} { \hat 2}}    = {\bf R}_{{ \hat 2} { \hat 2}}  
                                                                            =   R_{{ \hat 0} { \hat 3} { \hat 0} { \hat 3} } 
                                                                           = {\bf R}_{{ \hat 3} { \hat 3}} =    \frac{M}{R} \, ,\\
            R_{{ \hat 2} { \hat 3} { \hat 2} { \hat 3}}& = {\bf R}_{{ \hat 4} { \hat 4}} =   \frac{2 M}{R}  \, , \ \
            R_{{ \hat 3 }{ \hat 1} { \hat 3} { \hat 1}}   = {\bf R}_{{ \hat 5} { \hat 5}} =  R_{{ \hat 1} { \hat 2} { \hat 1} { \hat  2}}    
                                                                           = {\bf R}_{{ \hat 6} { \hat 6}} =   - \frac{M}{R} \, .
           \nonumber                                                            
            \end{align}            
 Consequently, the curvature matrix ${\bf R_{AB}}$ is diagonal with eigenvalues
            \begin{align}
              \lambda_{1}   &  =  - \lambda_{4}  = - \frac{2 M}{R}  \ ,  \ \
              \lambda_{2}        =    \lambda_{3}   =     - \lambda_{5} = - \lambda_{6}  =  \frac{M}{R} \ ,
              \label{eq: eigenvs_of_R_inN_V2}
                \end{align}
satisfying the relationship
             \begin{align}
              \sum_{i=1}^{6}\lambda_{i} = 0 \ ,
               \end{align}        
which agrees with equation (\ref{poi}) .

\section{Nearly Newtonian space-time to quadratic order in Fermi-normal coordinates}\label{sect:C}
This section aims to calculate the metric corresponding to the nearly Newtonian space-time to quadratic order 
in Fermi-normal coordinates surrounding a radial geodesic.  We start by considering the nearly Newtonian metric in the  
orthonormal tetrad previously discussed  in Equations (\ref{eq:NewtApp})  and,  (\ref{eq:one-form-a}) and (\ref{eq:one-form-b}), 
i.e.,
     \begin{align}
     {\cal G} =   -  d \mathcal{T} \otimes d \mathcal{T}  +  d \mathcal{ R } \otimes d \mathcal{ R }  
                      +  d\mathcal{\theta} \otimes  d\mathcal{\theta} 
                      +  d \mathcal{\varphi} \otimes d \mathcal{\varphi}  \ .
     \label{eq:NewtAppRect}                                                                                    
      \end{align}
To find the equations of a radial geodesic $\mathcal{T}(t)$,  $\mathcal{R}(t)$ with $\mathcal{\theta}$ and $\varphi$ 
constants, one may replace the geodesic equations with two first integrals:
           \begin{align}
             - \mathcal{T}'^{2}    +  \mathcal{R}'^{2}   & = -1 \ ,  \\
              - \mathcal{T}' & = - k \ ,
                     \end{align} 
the normalization of proper time and a dimensionless energy parameter $(-k)$, respectively.   Here,  `` $'$ '' denotes derivate 
with respect to the parameter $t$. By combining these  equations,  we get
           \begin{align}
           \mathcal{R}'^{2}  =k^2  -1 \ ,
            \end{align}
and, consequently, after renaming the constant $\alpha^2 \equiv {(k^2  -1)^{-1}}$ , we receive the geodesic equation given by
           \begin{align}
            dt^2 = \alpha^2 d \mathcal{R}^{2} \ .
            \end{align}
Finally, after using  the relations given in (\ref{eq:one-form-a}) and (\ref{eq:one-form-b}), we reach
           \begin{align}
            dt^2 =  \alpha^2 (1 -2 \phi)d{R}^{2} \ .
            \label{eq:NewtonGeodesic}
            \end{align}
This expression allows us to get a geodesic explicitly after specifying a Newtonian potential.  Along with this work, we will 
consider only central Newtonian potentials; therefore, this result indicates that one may take $R$ in place of proper time 
$t$  to identify points on this geodesic.

So far, we have studied a nearly Newtonian approach corresponding to a static and spherical gravitational source and have 
chosen a geodesic in that space-time. Usually, the difficult part of the subsequent calculations in resolving the Fermi coordinates 
is the determination of the orthonormal basis parallel transported on the reference geodesic.  However, the calculations are simple 
since we have expressed the space-time on the orthonormal basis as given in Equations (\ref{eq:one-form-a}) and (\ref{eq:one-form-b}).
 On this basis, the geodesic looks like a right line, and the transformation between the Fermi-normal coordinates 
$x^{\alpha} \equiv (t, x, y, z) $ and the arbitrary coordinates are then
                      $x^{\tilde \alpha} \equiv ( \mathcal{T}, \mathcal{R}, \mathcal{\theta}, \varphi)$ 
                     \begin{align}
                       \eta_{\alpha \beta} =  ({\mathbf e}_{\alpha})^{\hat \mu}  ({\mathbf e}_{\beta})^{\hat \nu}
                                                          \eta_{\hat \mu \hat \nu} \ ,
                        \end{align}
with                         
                      \begin{align}
                       \left( {\mathbf e}_{0},  {\mathbf e}_{1} ,  {\mathbf e}_{2} ,  {\mathbf e}_{3} \right)  \equiv
                       \left(\frac{\partial \mathcal{} }{ \partial t},  \frac{\partial \mathcal{} }{ \partial x}  ,  
                        \frac{\partial \mathcal{} }{ \partial y}  , \frac{\partial \mathcal{} }{ \partial z}  \right){\bigg  |}_{\gamma} \ .
                       \end{align}
Hence, the non-zero components of this basis in terms of the orthonormal tetrad previously discussed can be derived from             
                                          \begin{align}
                      {\mathbf e}_{0}  & \equiv \frac{\partial \mathcal{} }{ \partial t} {\bigg  |}_{\gamma} 
                                              = \mathcal{T}' \frac{\partial \mathcal{} }{ \partial\mathcal{T}} 
                                              + \mathcal{R}' \frac{\partial \mathcal{} }{ \partial\mathcal{R}} \ ,\\
                      {\mathbf e}_{1} & \equiv \frac{\partial \mathcal{} }{ \partial x}{\bigg  |}_{\gamma} 
                                              = {R}' \frac{\partial \mathcal{} }{ \partial\mathcal{T}} 
                                              + \mathcal{T}' \frac{\partial \mathcal{} }{ \partial\mathcal{R}} \ ,\\
                      {\mathbf e}_{2} & \equiv\frac{\partial \mathcal{} }{ \partial y} {\bigg  |}_{\gamma} 
                                              =   \frac{\partial \mathcal{} }{ \partial\mathcal{\theta}}   \ ,\\
                      {\mathbf e}_{3} & \equiv \frac{\partial \mathcal{} }{ \partial z} {\bigg  |}_{\gamma} 
                                              =  \frac{\partial \mathcal{} }{ \partial\varphi} \ .
                       \end{align}
After determining the orthonormal basis, the next step in expressing the metric in the Fermi-normal coordinates is to project the curvature 
tensor components given by equation  (\ref{eq:CurvatureSpherCoord})  in this basis. For this purpose, we use the transformation formula given by
                     \begin{align}
                      R_{\alpha \beta \mu \nu} = ({\mathbf e}_{\alpha})^{\hat \alpha}  ({\mathbf e}_{\beta})^{\hat \beta}
                                                                 ({\mathbf e}_{\mu})^{\hat \mu}  ({\mathbf e}_{\nu})^{\hat \nu}
                                                                 R_{ \hat{\alpha} \hat{\beta} \hat{\mu} \hat{\nu} } \ .
                      \end{align}       
Hence, the non-zero components of the Riemann curvature tensor are:   
                      \begin{align}
                       R_{0101}  & =  R_{\hat{0}\hat{1}\hat{0}\hat{1}}  =  \phi_{RR} \ ,  \\
                       R_{0303} & =  R_{0202}  =  { \mathcal{T}'}^2  R_{\hat{0}\hat{2}\hat{0}\hat{2}}  
                                           + { \mathcal{R}'}^2  R_{\hat{1}\hat{2}\hat{1}\hat{2}} 
                                           = k^2 \nabla^2\phi - (\phi_{RR} + \frac{1}{R} \phi_{R}) \ ,  \\                 
                       R_{2323}  & =    R_{\hat{2}\hat{3}\hat{2}\hat{3}}    = \frac{2}{R} \phi_{R}     \ ,  \\       
                       R_{1313}  & =  R_{1212}  =  { \mathcal{R}'}^2  R_{\hat{0}\hat{2}\hat{0}\hat{2}}  
                                         + { \mathcal{T}'}^2  R_{\hat{1}\hat{2}\hat{1}\hat{2}}
                                         = k^2 \nabla^2\phi -  \frac{1}{R} \phi_{R} \ .               
                         \end{align}      
Finally, after introducing the last components of the Riemann curvature tensor in equations (\ref{eq:F-N-Ma}),
 (\ref{eq:F-N-Mb}) and (\ref{eq:F-N-Mc}), we get the Fermi-normal metric corresponding to the Newtonian approximation,                      
                  \begin{align}   
                      \mathcal{G}  = &  - \left[ 1  - \phi_{RR} x^2 +  \left(\phi_{RR} 
                                                  + \frac{1}{R} \phi_{R} - k^2\nabla^2 \phi\right)(y^2 + z^2)  \right]  dt \otimes dt \nonumber\\
                                              &   - \frac{1}{3R} \phi_{R} \big[ ( ydx - xdy) \otimes ( ydx - xdy)
                                                   + ( xdz - zdx) \otimes ( xdz - zdx)  
                                                \nonumber \\ 
                                                & - 2 ( ydz - zdy) \otimes ( ydz - zdy) \big]   +  dx \otimes dx + dy \otimes dy + dz \otimes dz 
                                                \nonumber\\
                                                &  + \frac{k^2}{3R} \nabla^2\phi \big[ ( ydx - xdy) \otimes ( ydx - xdy) 
                                                   + ( xdz - zdx) \otimes ( xdz - zdx)\big]     
                       \label{eq:Fermi-NewtonMetric0}                                                                                                                         
                       \end{align}             
By introducing spherical coordinates $r, \theta, \varphi$  related to the rectangular coordinates $x, y, z$ by the standard formulas and taking 
the $x$ direction as the polar axis, i.e., 
                                                       $$ x  =  r \cos\theta \ , \quad
                                                            y  =  r \sin\theta \cos\varphi\ , \quad
                                                            z  =  r \sin\theta\sin\varphi \ ,
                                                       $$
the Fermi-normal metric corresponding to the nearly Newtonian space-time reads                              
                     \begin{align}   
                      \mathcal{G}  = & -  \Big[ 1 - \big( 2 \phi_{RR} + \frac{1}{R}\phi_{R}  - k^2\nabla^2\phi\big)r^2 \cos^2\theta 
                                                 + \big(  \phi_{RR} + \frac{1}{R}\phi_{R}  - k^2\nabla^2\phi\big)r^2  \Big]  dt \otimes dt  
                                                \nonumber\\
                                              &   +  dr \otimes dr   + 
                                              \Big[  1 - \frac{r^2}{3R} \phi_{R} + \frac{k^2}{3}r^2\nabla^2\phi    \Big] r^2 d\theta \otimes d\theta 
                                                 + \Big[  1 - \frac{r^2}{3R} \phi_{R} ( 3\cos^2\theta - 1) 
                                                 \nonumber\\
                                                & +\frac{r^2}{3R^2} \phi_R
                                                 + \frac{k^2}{3}r^2 \cos^2\phi \nabla^2\phi \Big] r^2 \sin^2\theta d\varphi  \otimes d\varphi \ .   
                         \label{eq:Fermi-NewtonMetric}                                      
                          \end{align}                      
To illustrate the Fermi-normal metric corresponding to the nearly Newtonian space-time,  let us consider a spherically symmetric solution. 
Then, the exterior field should correspond to that of a sphere described by a solution of the Laplace equation with    
                            \begin{align}                            
                             \rho = 0 \ , \quad   \phi= - \frac{M}{R} \ ,  
                              \end{align} 
where $M$ is a constant. It is then straightforward to calculate the Fermi-normal metric, which turns out to be                                                             
                     \begin{align}   
                      \mathcal{G}   = & - \left[1 + \mu (3\cos^2\theta -1) \right]dt \otimes dt  + dr \otimes dr 
                                                  +  \left(1 - \frac{\mu}{3}\right) r^2 d\theta \otimes d\theta 
                                                    \nonumber\\
                                               & + \left[1 - \mu (3\cos^2\theta -1) + \frac{\mu}{3}\right]r^2 \sin^2\theta d\varphi \otimes d\varphi \ ,
                                               \qquad
                              \mu    \equiv  \frac{Mr^2}{R^3} \ .
                           \label{eq:Fermi-NewtonMetricR}
                           \end{align} 
Consistently with this   metric, the geodesic (\ref{eq:NewtonGeodesic})  takes the particular form
           \begin{align}
            dt^2 =  \alpha^2 \left(1 + \frac{2M}{R}\right)d{R}^{2} \ . 
            \label{eq:PunctualNewtonGeodesic}
            \end{align}                  
In  Equation (\ref{eq:Fermi-NewtonMetric}),  we observe that performing a quadratic order of the nearly Newtonian gravitational 
field breaks the spherical symmetry if Fermi-Normal coordinates are used. In fact, even if the Newtonian potential has no angular
 dependence, the Fermi-normal metric corresponding to the Newtonian approximation contains mixed terms, implying functionality 
 in terms of angular variables. The reason for that angular dependence follows from the construction of the metric tensor in terms 
 of the Riemann curvature tensor. Another critical point to discuss in this part has to do with the radial coordinate $R$. 
 Attending  Equation (\ref{eq:PunctualNewtonGeodesic}), we observe that in Equation (\ref{eq:Fermi-NewtonMetricR}) $R$ must 
 be considered a function of $t$ or to take $R$ as the time coordinate.       
 The chief conclusion about all of this is that the description of the nearly Newtonian gravitational field on the basis of the 
 Fermi-normal coordinates, in a quadratic order of approximation in general relativity, exhibits different symmetry to that gravitational 
 field in the Newton theory employed to perform that approximation.

 In this section, we studied an example consisting of a Newtonian potential of a punctual mass. On the basis of the Fermi-normal coordinates, 
 in the quadratic order of approximation in general relativity, this corresponds to the potential of a no-static source with no spherical symmetry. 
 A similar result is obtained by performing a description of the Schwarszchild on the basis of the Fermi-normal coordinates in the quadratic 
 order of approximation. In this case, we obtain the space-time near a no-static source with deformation parameters different from zero.                     
We finalize this section by calculating the shape of a sphere whose center of mass coincides with the coordinates of 
some point of the central geodesic. To do so, we define a sphere $\Sigma$ as the surface formed by all points a proper fixed distance $r$ measured 
orthogonally from such a point. For the coordinates of Equation (\ref{eq:Fermi-NewtonMetric}) this is the surface $t=$ constant, $r=$ constant, whose metric is,
                                       \begin{align}
                                       \mathcal{G}{\Big |}_\Sigma =   \left(1 - \frac{\mu}{3}\right) r^2 d\theta \otimes d\theta 
                                                                               + \left[1 - \mu (3\cos\theta -1) 
                                                                               + \frac{\mu}{3}\right]r^2 \sin^2\theta d\varphi \otimes d\varphi \ .
                                        \end{align}                   
The length  of the great circle $ \varphi=$ constant over the poles of this sphere is                                       
                      \begin{align} 
                      L_{\text {Poles}} = \int^{2\pi}_0 \left(1 - \frac{\mu}{3}\right)^{1/2} r d\theta  \approx  2\pi r \left(1 - \frac{\mu}{6}\right) \ .
                       \end{align}  
The circumference of   the  equator, $\theta = \pi/2$, is 
                      \begin{align} 
                      L_{\text {Equator}} = \int^{2\pi}_0 \left[1 - \frac{\mu}{3} (3\cos\theta -1)  + \frac{\mu}{3}\right]^{1/2} 
                                                                          r \sin\theta d\varphi \mid_{\theta =\pi/2} \;\;
                        \approx  2\pi r \left(1 + \frac{\mu}{3}\right) \ .
                       \end{align}  
Consequently, the distortion of the shape $\eta$ of the sphere is given by
                       \begin{align} 
                         \eta \equiv {\bigg |}\frac{  L_{\text {Poles}} -  L_{\text {Equator}} }{  L_{\text {Poles}}  
                                                                                           +   L_{\text {Equator}} }{\bigg |}
                                = \frac{\mu}{4} =\frac{M r^2}{4 R^3} \ .
                        \end{align}                          
This result indicates that a sphere $r$=constant is a surface shaped like a prolate spheroid.

\section{CONCLUDING REMARKS}\label{sect:D}
In this work, we find the metric corresponding to a static gravitational source for a weak-field approximation to 
quadratic order in fermi-normal coordinates surrounding a radial geodesic. We conclude that in the lowest order, 
the effects of a Schwarzschild gravitational field measured locally by a freely falling observer equals the measured
 by an observer in similar conditions in the presence of a Newtonian approximation of gravitation.

To simplify the calculations in this work, we introduced a general metric, expressed it in an orthonormal tetrad, 
and used the Cartan formalism of differential forms. Incidentally, we cast out the eigenvalues of the Riemann tensor 
corresponding to the nearly Newtonian gravitation. It is possible to directly compare these eigenvalues and the corresponding
 to the nearly Newtonian gravitation in its quadratic approximation. This comparison reveals that the metric corresponds 
 to a non-static space-time in its quadratic approximation.

The results obtained in this work indicate the possibility of studying the nearly Newtonian gravitation in its quadratic
 approximation in other theories and scenarios, for instance, the problem of two gravitational sources, the interaction 
 between gravitational sources in Einstein-Maxwell gravity, etcetera. The discussion presented here can be generalized 
 to stationary environments and rotating observers. It can be used in practical problems to improve understanding of relativistic effects, 
 for example, the Sagnac effect. We expect to consider this problem in future works.

 
 \appendix                                               
\section{Fermi normal coordinates for the Schwarzschild solution}\label{sect:E}      
In this section, we compute the metric corresponding to the Schwarzschild solution to quadratic order in Fermi-normal 
coordinates surrounding a radial geodesic. The Schwarzschild metric in a not-rotating local frame of reference in spherical 
coordinates $y^{a} = (T, R, \Theta, \Phi)$ admits the form 
     \begin{align}
     {\cal G} =   - X dT \otimes dT  +  \frac{1}{X}dR \otimes dR  
                  + R^2 \left(d\Theta \otimes d\Theta  + \sin^2 \Theta d\Phi \otimes d\Phi \right),\ 
                   X \equiv 1 - \frac{2M}{R} \ .
      \label{eq:SchwarzMetric}          
      \end{align}
 To use the Cartan formalism, we introduce the orthonormal tetrad  
      \begin{align}
       \vartheta^{\hat{0}}  & = X^{1/2} dT \ \equiv d \mathcal{T},\quad
         \vartheta^{\hat{1}} = X^{-1/2} dR\   \equiv d \mathcal{ R },  \label{eq:one-formSchwarz-a}\\
              \vartheta^{\hat{2}}  &=  R \, d\Theta \equiv  d\mathcal{\theta} \ ,\quad
              \vartheta^{\hat{3}}  =  R \sin\Theta \, d\Phi \equiv d \mathcal{\varphi}\ ,
              \label{eq:one-formSchwarz-b}
               \end{align}
which leads to the metric in the rectangular form 
     \begin{align}
     {\cal G} =   -  d \mathcal{T} \otimes d \mathcal{T}  +  d \mathcal{ R } \otimes d \mathcal{ R }  
                      +  d\mathcal{\theta} \otimes  d\mathcal{\theta} 
                      +  d \mathcal{\varphi} \otimes d \mathcal{\varphi}  \ ,
     \label{eq:SchwarzschildRect}                                                                                    
      \end{align}
and the connection 1-form  
          \begin{align}
           \omega^{\hat 1}_{\  \hat 0}  & = -\frac{M}{R^3} X^{-1/2}   \vartheta^{\hat 0},  \quad  
	   \omega^{\hat 2}_{\ \hat 3}  = -\frac{1}{R} \cot{\Theta} \vartheta^{ \hat 3}, \\
            \omega^{ \hat 1}_{\ \hat 2}   & = -\frac{1}{R} X^{1/2} \vartheta^{ \hat 2}, \quad
             \omega^{ \hat 1}_{\ \hat 3}   = - \frac{1}{R} X^{1/2} \vartheta^{ \hat 3} \ .
               \end{align}
It then follows that the only non-zero components  of  the curvature  tensor are 
          \begin{align}
            R_{{ \hat 0} { \hat 1} { \hat 0} { \hat 1}} & = {\bf R}_{{ \hat 1} { \hat 1}} =   - \frac{2 M}{R} \, , \ \
            R_{{ \hat 0} { \hat 2} { \hat 0} { \hat 2}}    = {\bf R}_{{ \hat 2} { \hat 2}}  
                                                                            =   R_{{ \hat 0} { \hat 3} { \hat 0} { \hat 3} } 
                                                                           = {\bf R}_{{ \hat 3} { \hat 3}} =    \frac{M}{R} \, ,\\
            R_{{ \hat 2} { \hat 3} { \hat 2} { \hat 3}}& = {\bf R}_{{ \hat 4} { \hat 4}} =   \frac{2 M}{R}  \, , \ \
            R_{{ \hat 3 }{ \hat 1} { \hat 3} { \hat 1}}   = {\bf R}_{{ \hat 5} { \hat 5}} =  R_{{ \hat 1} { \hat 2} { \hat 1} { \hat  2}}    
                                                                           = {\bf R}_{{ \hat 6} { \hat 6}} =   - \frac{M}{R} \, .
            \end{align}
 Consequently, the curvature matrix ${\bf R_{AB}}$ is diagonal with eigenvalues
            \begin{align}
              \lambda_{1}   &  =   - \frac{2 M}{R}  \ ,  \ \
              \lambda_{2}        =    \lambda_{3}   =      \frac{M}{R} \ , \\ 
              \lambda_{4}   &  =    \frac{2 M}{R} \ , \ \
              \lambda_{5}       =     \lambda_{6}  =     - \frac{ M}{R}   \ ,
              \label{eq: eigenvs_of_R_inN_V2}
                \end{align}
which satisfies the relationship
             \begin{align}
              \sum_{i=1}^{6}\lambda_{i} = 0 \ ,
               \end{align}             
in consistency with the fact that the Schwarzschild solution is a vacuum solution of the Einstein equations.

To find the equations of a radial geodesic $\mathcal{T}(t)$,  $\mathcal{R}(t)$ with $\mathcal{\theta}$ and $\varphi$ 
constants, one may replace the geodesic equations with two first integrals:
           \begin{align}
             - \mathcal{T}'^{2}+  \mathcal{R}'^{2}  & = -1 \ , \\
              - \mathcal{T}' & = - k \ ,
                     \end{align} 
the normalization of proper time and a dimensionless energy parameter $(-k)$, respectively.
Here,  `` $'$ '' denotes derivate with respect to the parameter $t$. After combining these  equations,  we get
           \begin{align}
           \mathcal{R}'^{2}  = k^2  -1 \ ,
            \end{align}
and, consequently, after renaming the constant  $\alpha^2 = {(k^2  -1)^{-1}} $ we receive the geodesic equation given by
           \begin{align}
            dt^2 = \alpha^2 d \mathcal{R}^{2} \ .
             \label{eq:geodSchwarz1}
            \end{align}
Finally, after using the relations given in (\ref{eq:one-formSchwarz-a}) and (\ref{eq:one-formSchwarz-b}), we reach
          \begin{align}
            dt^2 =  \frac{\alpha^2}{1-2M/R}d{R}^{2}  \ . 
            \label{eq:geodSchwarz2}
            \end{align}
This expression allows us to get a radial geodesic in the Schwarzschild background. As we can see, because of
 its functional dependence, one may take $R$ in place of proper time $t$ to identify points on this geodesic.   As we 
 can see, similarly to the early Newtonian approach, the calculations are simple since we have expressed the space-time
 on the orthonormal basis as given in Equations (\ref{eq:one-formSchwarz-a}) and (\ref{eq:one-formSchwarz-b})  . On this basis, 
  the geodesic looks like a right  line, and the transformation between the Fermi-normal coordinates 
 $x^{\alpha} \equiv (t, x, y, z) $ and the arbitrary coordinates are then
                      $x^{\tilde \alpha} \equiv ( \mathcal{T}, \mathcal{R}, \mathcal{\theta}, \varphi)$ 
                     \begin{align}
                       \eta_{\alpha \beta} =  ({\mathbf e}_{\alpha})^{\hat \mu}  ({\mathbf e}_{\beta})^{\hat \nu}
                                                          \eta_{\hat \mu \hat \nu} \ 
                        \end{align}
with                         
                      \begin{align}
                       \left( {\mathbf e}_{0},  {\mathbf e}_{1} ,  {\mathbf e}_{2} ,  {\mathbf e}_{3} \right)  \equiv
                       \left(\frac{\partial \mathcal{} }{ \partial t},  \frac{\partial \mathcal{} }{ \partial x}  ,  
                        \frac{\partial \mathcal{} }{ \partial y}  , \frac{\partial \mathcal{} }{ \partial z}  \right){\bigg  |}_{\gamma} \ .
                       \end{align}
Hence, the non-zero components of this basis in terms of the orthonormal tetrad previously discussed can be derived from             
                                          \begin{align}
                      {\mathbf e}_{0}  & \equiv \frac{\partial \mathcal{} }{ \partial t} {\bigg  |}_{\gamma} 
                                              = \mathcal{T}' \frac{\partial \mathcal{} }{ \partial\mathcal{T}} 
                                              + \mathcal{R}' \frac{\partial \mathcal{} }{ \partial\mathcal{R}} \ ,\\
                      {\mathbf e}_{1} & \equiv \frac{\partial \mathcal{} }{ \partial x}{\Big  |}_{\gamma} 
                                              = {R}' \frac{\partial \mathcal{} }{ \partial\mathcal{T}} 
                                              + \mathcal{T}' \frac{\partial \mathcal{} }{ \partial\mathcal{R}} \ ,\\
                      {\mathbf e}_{2} & \equiv\frac{\partial \mathcal{} }{ \partial y} {\Big  |}_{\gamma} 
                                              =   \frac{\partial \mathcal{} }{ \partial\mathcal{\theta}}   \ ,\\
                      {\mathbf e}_{3} & \equiv \frac{\partial \mathcal{} }{ \partial z} {\Big  |}_{\gamma} 
                                              =  \frac{\partial \mathcal{} }{ \partial\varphi} \ .
                       \end{align}

After determining the orthonormal basis, the next step in expressing the metric in the Fermi-normal coordinates is to project
 the curvature tensor components given by equation  (\ref{eq:CurvatureSpherCoord})  in this basis. For this purpose, 
 we use the transformation formula given by
                     \begin{align}
                      R_{\alpha \beta \mu \nu} = ({\mathbf e}_{\alpha})^{\hat \alpha}  ({\mathbf e}_{\beta})^{\hat \beta}
                                                                 ({\mathbf e}_{\mu})^{\hat \mu}  ({\mathbf e}_{\nu})^{\hat \nu}
                                                                 R_{ \hat{\alpha} \hat{\beta} \hat{\mu} \hat{\nu} } \ .
                      \end{align}       
Hence, the non-zero components of the Riemann curvature tensor are:   
                      \begin{align}
                       R_{0101}  & =  R_{\hat{0}\hat{1}\hat{0}\hat{1}}  =  -\frac{2M}{R^3} \ ,  \\
                       R_{0303} & =  R_{0202}  =  { \mathcal{T}'}^2  R_{\hat{0}\hat{2}\hat{0}\hat{2}}  
                                           + { \mathcal{R}'}^2  R_{\hat{1}\hat{2}\hat{1}\hat{2}} 
                                           = \frac{M}{R^3} \ , \\                 
                       R_{2323}  & =    R_{\hat{2}\hat{3}\hat{2}\hat{3}}    = \frac{M}{R^3}    \ ,  \\       
                       R_{1313}  & =  R_{1212}  =  { \mathcal{R}'}^2  R_{\hat{0}\hat{2}\hat{0}\hat{2}}  
                                         + { \mathcal{T}'}^2  R_{\hat{1}\hat{2}\hat{1}\hat{2}}
                                         =- \frac{M}{R^3} \ .               
                         \end{align}   
Finally, after introducing the last components of the Riemann curvature tensor in equations (\ref{eq:F-N-Ma}),
 (\ref{eq:F-N-Mb}) and (\ref{eq:F-N-Mc}), we get the Fermi-normal metric corresponding to the Schwarzschild space-time,   
                     \begin{align}   
                      \mathcal{G}  = &  -    \left[ 1  - \frac{M}{R^3}(y^2 + z^2 -2x^2)  \right]  dt \otimes dt  \nonumber \\
                                              &  +    \frac{2M}{3R^3}  \left[ xy  dx \otimes dy  + xz dx \otimes dz  - 2yz dy \otimes dz \right]     
                                              \nonumber\\
                                               & +    \left[   1  - \frac{M}{3R^3} (y^2 + z^2)  \right]  dx \otimes dx     \nonumber\\
                                               & +    \left[   1  - \frac{M}{3R^3} (x^2 -2 z^2)  \right]  dy \otimes dy      \nonumber\\
                                               &   +    \left[   1  - \frac{M}{3R^3} (x^2 - 2 y^2)  \right]  dz \otimes dz \ .  
                                   \end{align}             
By introducing spherical coordinates $r, \theta, \varphi$  related to the rectangular coordinates $x, y, z$ by the standard formulas 
and taking the $x$ direction as the polar axis, i.e. , 
                                                       $$ x  =  r \cos\theta \ , \quad
                                                            y  =  r \sin\theta \cos\varphi\ , \quad
                                                            z  =  r \sin\theta\sin\varphi \ ,
                                                       $$
the Fermi-normal metric corresponding to the Schwarzschild space-time reads                                                                                                          
                     \begin{align}   
                      \mathcal{G}   = & - \left[1 + \mu (3\cos^2\theta -1) \right]dt \otimes dt  + dr \otimes dr 
                                                  +  \left(1 - \frac{\mu}{3}\right) r^2 d\theta \otimes d\theta 
                                                    \nonumber\\
                                               & + \left[1 - \mu (3\cos^2\theta -1) + \frac{\mu}{3}\right]r^2 \sin^2\theta d\varphi \otimes d\varphi \ ,
                                               \qquad
                              \mu    \equiv  \frac{Mr^2}{R^3} \ .
                           \label{eq:Fermi-SchwarzschildMetric}
                           \end{align}

%


\end{document}